\documentclass[
    final, sort&compress          % use final for the camera ready runs
   ]
  {aipproc}

\layoutstyle{8x11double}

\usepackage{amssymb}
\usepackage[intlimits]{amsmath}
\usepackage{amsfonts}
\usepackage{dsfont}
\usepackage[usenames,dvipsnames]{color}
\usepackage{colortbl}

\usepackage{bm}
\usepackage{multirow}
\usepackage{bbm}
\usepackage{slashed}
\usepackage{textcomp}

        \usepackage[vcentermath]{youngtab}

      \newcommand{\mA}{\mathsf{A}}
      \newcommand{\mS}{\mathsf{S}}
      \newcommand{\mD}{\mathsf{D}}
      \newcommand{\mT}{\mathsf{T}}

      \newcommand{\vect}[1]{\bm{#1}}

\begin{document}

\title{The muon g-2: Dyson-Schwinger status on \\ hadronic light-by-light scattering}


\author{Gernot Eichmann}{}
\author{Christian S. Fischer}{}
\author{Walter Heupel}{}
\author{Richard Williams}{
  address={Institut f\"ur Theoretische Physik, Justus-Liebig--Universit\"at Giessen, 35392 Giessen, Germany}
}

\begin{abstract}
    We give a status report on the hadronic light-by-light scattering contribution to the muon's anomalous magnetic moment from the Dyson-Schwinger approach.
    We discuss novel, model-independent properties of the light-by-light amplitude:
    we give its covariant decomposition in view of electromagnetic gauge invariance and Bose symmetry,
    and we identify the relevant kinematic regions that are probed under the integral. 
    The decomposition of the amplitude at the quark level and the importance of its various diagrams are discussed and related to model approaches.

\end{abstract}

\maketitle

%%%%%%%%%%%%%%%%%%%%%%%%%%%%%%%%%%%%%%%%%%%%
%% MAINMATTER
%%%%%%%%%%%%%%%%%%%%%%%%%%%%%%%%%%%%%%%%%%%%

\subsection{Introduction}

     The puzzle of the muon anomalous magnetic moment $a_\mu=(g_\mu-2)/2$ continues to pose
     a challenge across various communities in particle physics.
     The current $3\sigma$ discrepancy between the experimental value
     and the Standard Model prediction (both are quoted in Table~\ref{tab:g-2})
     has spurred developments in QED, QCD and physics beyond the Standard Model,
     see~\cite{Jegerlehner:2009ry} for a comprehensive review.
     In addition, new measurements at Fermilab~\cite{LeeRoberts:2011zz} and JPARC~\cite{Iinuma:2011zz}
     have been proposed to further improve upon the present experimental precision.

     The magnetic moment of a lepton $l=e,\mu$ is encoded in its electromagnetic current:\footnote{We
         work in a Euclidean metric with the following
         replacement rules for Lorentz vectors $a^\mu$, tensors $T^{\mu\nu}$, and $\gamma-$matrices:
         \begin{equation}\label{rules}
             a^\mu_\text{E} = \left[\begin{array}{c}\vect{a}\\ia_0\end{array}\right], \quad
             \gamma^\mu_\text{E} = \left[\begin{array}{c}-i\vect{\gamma}\\\gamma_0\end{array}\right], \quad
             T^{\mu\nu}_\text{E} = \left[\begin{array}{cc} T^{ij} & iT^{i0} \\ iT^{0i} & -T^{00}\end{array}\right].
         \end{equation}
         'E' stands for Euclidean and the absence of a subscript for Minkowski conventions.
         This entails $a_\text{E} \cdot b_\text{E} = -a\cdot b$ for scalar products, $\slashed{a}_\text{E} = i\slashed{a}$,
         $g^{\mu\nu}_\text{E} = -\delta^{\mu\nu}$, and $\{ \gamma^\mu_\text{E}, \gamma^\nu_\text{E} \} = 2\delta^{\mu\nu}$.
         We drop the label 'E'.}
     \begin{equation}
         J_l^\mu = \bar{u}(p')\left[ F_1^l(Q^2)\,\gamma^\mu - \frac{F_2^l(Q^2)}{2m_l}\,\sigma^{\mu\nu} Q^\nu \right] u(p)\,,
     \end{equation}
     where $\sigma^{\mu\nu} = -\tfrac{i}{2} \,[\gamma^\mu,\gamma^\nu]$.
     The anomalous magnetic moment is the  Pauli form factor at zero momentum transfer: $a_l=F_2^l(0)$.
     Its leading contribution is Schwinger's result~\cite{Schwinger:1948iu} for the one-loop dressing of the lepton-photon vertex,
     $a_l = \alpha/(2\pi) + \mathcal{O}(\alpha^2)$.
     The majority of corrections come from QED and are known up to $\mathcal{O}(\alpha^5)$, with
     uncertainties from higher orders in perturbation theory already below the measurement error~\cite{Aoyama:2012wk}.
     Further electroweak and QCD corrections are strongly suppressed.
     They enter with magnitude $10^{-8}$ for the muon but only $10^{-12}$ for the electron because they scale with the squared lepton mass. 
     Since the current experimental precision is of the order $10^{-10}$ for the muon and $10^{-12}$ for the electron,
     these corrections are compatible with
     the experimental error in the case of $a_e$, whereas for $a_\mu$ they are by two orders of magnitude larger.

        \renewcommand{\arraystretch}{1.3}

        \begin{table}[t]

                \begin{tabular}{  @{\;\;} l @{\;\;\;\;}        @{\;\;\;}r@{\;\;\;}   @{\;\;\;\;}r@{\;\;}   @{\;\;\;\;}l@{\;\;}     } \hline\noalign{\smallskip}

                    \textbf{Experiment}       & $11\,\,659\,\,208.9$   & $(6.3)$  &  \cite{Bennett:2004pv,Agashe:2014kda}                       \\  \noalign{\smallskip}\hline\noalign{\smallskip}
                    QED                       & $11\,\,658\,\,471.9$   & $(0.0)$  &  \cite{Aoyama:2012wk}   \\
                    Electroweak               &               $15.3$   & $(0.2)$  &  \cite{Jegerlehner:2009ry}   \\
                    Hadronic VP               &              $685.1$   & $(4.3)$  &  \cite{Hagiwara:2011af}  \\

                    Hadronic LbL              &               $11.6$   & $(3.9)$  &  \cite{Jegerlehner:2009ry}  \\  \noalign{\smallskip}\hline\noalign{\smallskip}
                    \textbf{Standard Model}   & $11\,\,659\,\,183.9$   & $(5.8)$  &    \\
                    \textbf{Difference}       &               $25.0$   & $(8.6)$  &   \\  \noalign{\smallskip}\hline\noalign{\smallskip}

%                    Hadronic LbL              &               $10.5$   & $(2.6)$  &  \cite{Prades:2009tw}  \\  \noalign{\smallskip}\hline\noalign{\smallskip}
%                    \textbf{Standard Model}   & $11\,\,659\,\,182.8$   & $(4.9)$  &    \\
%                    \textbf{Difference}       &               $26.1$   & $(8.0)$  &   \\  \noalign{\smallskip}\hline\noalign{\smallskip}

                \end{tabular}

               \caption{ SM contributions to $a_\mu$ in units of $10^{-10}$ (quoted only up to another digit).}
               \label{tab:g-2}

        \end{table}

        \renewcommand{\arraystretch}{1.0}

        \begin{figure}[t]
                    \includegraphics[scale=0.11]{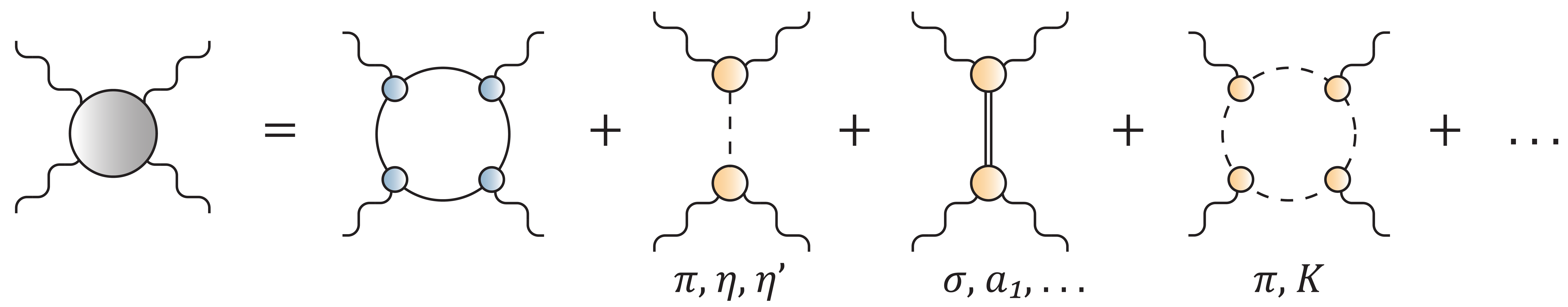}
                    \caption{Contributions to the LbL amplitude: quark loop, meson exchange diagrams and pion loops. }\label{fig:lbl-amplitude-models}
        \end{figure}

     The two types of QCD corrections are the hadronic vacuum polarization and hadronic light-by-light (LbL) scattering contributions.
     The former is the dominant correction and experimentally constrained from the total hadronic cross section in $e^+ e^-$ annihilation. 
     The LbL part emerges from the photon four-point function in Fig.~\ref{fig:lbl-amplitude-models} when three of the photons are coupled to the muon.
     Its contribution to $a_\mu$ is relatively small, but
     the fact that the uncertainty of the SM prediction is dominated by QCD effects warrants its closer inspection.
     The value in Table~\ref{tab:g-2} relies on a variety of models, see~\cite{Prades:2009tw,Jegerlehner:2009ry} for overviews.
     Current progress is also being made with dispersion relations~\cite{Colangelo:2014dfa,Colangelo:2014pva}, and lattice simulations are underway~\cite{Blum:2014oka}.
     In this work we will highlight recent developments in the Dyson-Schwinger equation (DSE) approach
     in determining the LbL contribution to the muon anomalous magnetic moment~\cite{Fischer:2010iz,Goecke:2010if,Goecke:2012qm}.

     The paper is organized as follows. First we give a brief overview of select model approaches to hadronic LbL.
     This is followed by summarizing the present status of the \textit{quark loop} calculation in the Dyson-Schwinger approach
     and its role in view of a consistent, gauge-invariant description of the LbL amplitude.
     Finally, we present a general discussion of the photon four-point function based upon
     constraints from Bose symmetry and electromagnetic gauge invariance.

\subsection{Model results}

     Fig.~\ref{fig:lbl-amplitude-models} shows the dominant diagrams that contribute to the LbL scattering amplitude
     and have been investigated in the literature: the quark loop, hadronic
     (pseudoscalar, scalar, axialvector, etc.) exchange diagrams, and pseudoscalar loops.
     They have been studied in a variety of approaches such as the
     extended Nambu-Jona-Lasinio (ENJL) model~\cite{Bijnens:1995cc,Bijnens:1995xf},
     quark models~\cite{Dorokhov:2008pw,Dorokhov:2011zf,Greynat:2012ww,Dorokhov:2012qa},
     and hadronic models based on light meson dominance~\cite{Hayakawa:1995ps,Hayakawa:1997rq,Knecht:2001qf,Melnikov:2003xd,Pauk:2014rta};
     for overviews we refer to Refs.~\cite{Prades:2009tw,Jegerlehner:2009ry,Dorokhov:2014iva,Masjuan:2014rea,Knecht:2014sea}.
     Quoting only ballpark numbers, the various contributions to $a_\mu$ are (in units of $10^{-10}$):
     \begin{itemize}
     \item pseudoscalar exchange: $8 \dots 11$,
     \item scalar exchange: $-1$,
     \item axialvector exchange: $2$,
     \item pseudoscalar loops: $-2$.
     \end{itemize}
     These numbers are typical for some of the models:
     the largest contribution comes from pion exchange, whereas the other effects are smaller.

     Another question is whether the quark loop should be included as well.
     In principle, if the hadronic description were complete in the sense of a spectral representation, such a term would introduce potential double-counting.
     As we will discuss below, Fig.~\ref{fig:lbl-amplitude-models} can be also viewed from a different perspective:
     each diagram can be identified with a distinct contribution of a systematic quark-level decomposition based upon gauge invariance.

     The constituent-quark loop result for $a_\mu$ is known analytically from
     the electron loop that appears in the LbL contribution of QED~\cite{Laporta:1992pa,Greynat:2012ww}:
     \begin{equation}\label{ql-analytic}
       a_\mu^\text{(QL)} = N_c \left(\frac{\alpha}{\pi}\right)^3 \sum_f q_f^4 \left[ c_1\,R_f^2 - c_2\,R_f^4 + \dots \right].
     \end{equation}
     $R_f=m_\mu/m_f$ is here the ratio of muon and quark masses, and the charges of the quark flavors $f=u,d,s$ are denoted by $q_f$.
     Inserting a constituent-quark mass in the window $m \sim 240 \dots 280$~MeV yields values in the range $a_\mu^\text{(QL)} \sim 6 \dots 8 \times 10^{-10}$.

     In the ENJL model the quark loop result is improved by dressing also the quark-photon vertex.
     Upon summing up quark bubbles, the vertex acquires a transverse part which has a dynamical $\rho-$meson pole:
     \begin{equation}\label{qpv-njl}
        \Gamma^\mu = i\gamma^\mu - \frac{Q^2}{Q^2+m_V^2}\,i\gamma^\mu_T\,,
     \end{equation}
     where $\gamma^\mu_T = T^{\mu\nu}_Q \gamma^\nu$ and $T^{\mu\nu}_Q$ is the transverse projector
     \begin{equation}\label{trans-proj-0}
         T^{\mu\nu}_Q = \delta^{\mu\nu} - \frac{Q^\mu Q^\nu}{Q^2}\,.
     \end{equation}
     It turns out that this additional transverse piece suppresses the magnitude of the quark loop,
     with the result $a_\mu^\text{(QL)} \sim 2 \times 10^{-10}$~\cite{Bijnens:1995cc,Bijnens:1995xf}. Hence, the quark loop is again a small effect,
     and adding all diagrams in Fig.~\ref{fig:lbl-amplitude-models} yields the QCD prediction for the LbL contribution quoted in Table~\ref{tab:g-2}.

\subsection{DSE calculation of the quark loop}

      How can one improve upon these calculations?
      To begin with, the quark mass is not a constant
      because the dressed quark propagator has a nonperturbatively enhanced quark mass function:
      \begin{equation}
         S_0(p) = \frac{-i\slashed{p}+m}{p^2+m^2} \;\; \rightarrow \;\;  S(p) = \frac{1}{A}\,\frac{-i\slashed{p}+M}{p^2+M^2}\,,
      \end{equation}
      where $M(p^2)$ and $A(p^2)$ are momentum-dependent functions.
      At large spacelike momenta, $M(p^2)$ approaches the current-quark mass with a logarithmic falloff.
      At low momenta the mass function becomes large and can be interpreted as the constituent-quark mass scale;
      for light quarks, $M(p^2=0) \sim 300 \dots 400$ MeV.
      These values are typical results from DSE studies in Landau gauge, % which depend on the form of the quark-gluon vertex,
      but they are also consistent with lattice data~\cite{Bowman:2005vx,Kamleh:2007ud} as well as their qualitative momentum dependence.
      $A(p^2)$ depends on the renormalization condition but it is also enhanced at low momenta, usually with a factor $\sim 1.5 \dots 2$
      compared to its perturbative value.

      On the other hand, dressing the quark will also change the quark-photon vertex.
      As a consequence of electromagnetic gauge invariance, the vertex satisfies the Ward-Takahashi identity (WTI)
      \begin{equation}\label{qpv-wti}
          Q^\mu \,\Gamma^\mu (k,Q) = S^{-1}(k_+)-S^{-1}(k_-) \,,
      \end{equation}
      which depends on the dressed quark propagator.
      ($Q$ is here the photon momentum and $k=\tfrac{1}{2}(k_++k_-)$ is the average momentum of the incoming and outgoing quarks.)
      If the propagator is known,
      then also the quark-photon vertex is determined up to transverse parts:
      \begin{equation}\label{vertex:BC}
          \Gamma^\mu(k,Q) =   \big[ i\gamma^\mu\,\Sigma_A + 2 k^\mu (i\slashed{k}\, \Delta_A  + \Delta_B)\big] + i\Gamma^\mu_T(k,Q)\,.
      \end{equation}
      The bracket on the r.h.s. is the Ball-Chiu vertex that is sufficient to satisfy the electromagnetic WTI~\cite{Ball:1980ay}.
      It depends on sums and difference quotients of the propagator dressing functions:
      \begin{equation}
          \Sigma_F = \frac{F(k_+^2) + F(k_-^2)}{2}\,, \quad
          \Delta_F = \frac{F(k_+^2) - F(k_-^2)}{k_+^2-k_-^2}\,,
      \end{equation}
      where $F \in \{A,B\}$ and $B=AM$.
      The transverse part $\Gamma^\mu_T$ is not constrained by gauge invariance and must be determined dynamically. Its generic form is
      \begin{equation}
          \Gamma^\mu_T(k,Q) = f_1\,Q^2 \,\gamma^\mu_T + \dots + f_3\,\tfrac{i}{2}\,[\gamma^\mu,\slashed{Q}] + \dots\,,
      \end{equation}
      with six further tensor structures that depend not only on the photon momentum $Q$ but also on the quark momentum $k$ (see, e.g., Refs.~\cite{Skullerud:2002ge,Eichmann:2014qva} for the complete expression).
      In general, the dressing functions $f_i$ depend on all Lorentz invariants: $k^2$, $k\cdot Q$ and  $Q^2$.
      Note that for a tree-level propagator with $M=m$ and $A=1$, together with $f_1 = -1/(Q^2+m_V^2)$ and all other $f_i=0$, Eq.~\eqref{vertex:BC} reproduces
      the ENJL version in Eq.~\eqref{qpv-njl}, because $\Sigma_A=1$ and the difference quotients vanish.

      Therefore, in order to calculate the dressed quark loop numerically, one needs to know the
      dressing functions $M(p^2)$ and $A(p^2)$ of the quark propagator, together with the eight transverse
      dressings $f_i(k^2, k\cdot Q, Q^2)$ of the quark-photon vertex.
      Both can be obtained in the Dyson-Schwinger formalism, where
      the quark propagator is solved from its DSE and
      the quark-photon vertex from its inhomogeneous Bethe-Salpeter equation.
      The latter reproduces the form of Eq.~\eqref{vertex:BC}: it self-consistently generates the Ball-Chiu vertex
      plus a transverse part that contains timelike vector-meson poles~\cite{Maris:1999bh}.
      This is important in the context of hadronic form factors: when form factors are calculated microscopically,
      the Ball-Chiu part of the vertex is necessary to reproduce the hadron's charge at $Q^2=0$, whereas the transverse part introduces
      the timelike pole structure.
      The leading transverse component $f_1$ is similar to that in Eq.~\eqref{qpv-njl} except that it also falls off with $k^2$,
      see~\cite{Maris:1999bh,Goecke:2012qm,Eichmann:2014qva} for details.

      The DSE calculation of the dressed quark loop with these ingredients has been performed in a series of works~\cite{Fischer:2010iz,Goecke:2010if,Goecke:2012qm}.
      The current result for the muon anomalous magnetic moment is quoted in Ref.~\cite{Goecke:2012qm}; with four flavors, $a_\mu^{\text{(QL)}} = 10.7(2) \times 10^{-10}$.
      It is five times larger than the ENJL value and, given the same error estimate, would reduce the discrepancy in Table~\ref{tab:g-2} to $\sim 2\sigma$.
      How does it come about?

      For the sake of illustration we will again use only ballpark numbers; a detailed discussion is given in~\cite{Goecke:2012qm}.
      Suppose we start from a constituent-quark loop with a bare vertex, $M=200$ MeV and $A=1$.
      From Eq.~\eqref{ql-analytic} the result for two light quark flavors is $a_\mu \sim 10$ (in units of $10^{-10}$).
      This does not change if we dress the quark with the mass function $M(p^2)$ calculated from its DSE,
      which means that the integrated average mass function is $M \sim 200$ MeV.
      Including in addition the DSE result for $A(p^2)$ leads to
      a suppression because $A(p^2)$ appears in the denominator of the quark propagator $\Rightarrow a_\mu \sim 5$.
      That effect is compensated if we switch on the $\Sigma_A$ dressing of the $\gamma^\mu-$component of the vertex, which
      leads back to $a_\mu \sim 10$.
      Adding now the leading transverse part of the vertex with $f_1=-1/(Q^2+m_V^2)$ induces a large
      suppression similar to that in the ENJL model: $a_\mu \sim 4$.
      However, upon including the result for $f_1$ calculated from the Bethe-Salpeter equation, which naturally also includes the dependence on $k^2$, one arrives again at $a_\mu \sim 10$.
      Therefore, it is the additional quark momentum dependence of the vertex that `suppresses the suppression' found in the ENJL model.

          \begin{figure}[t]
                    \includegraphics[scale=0.105]{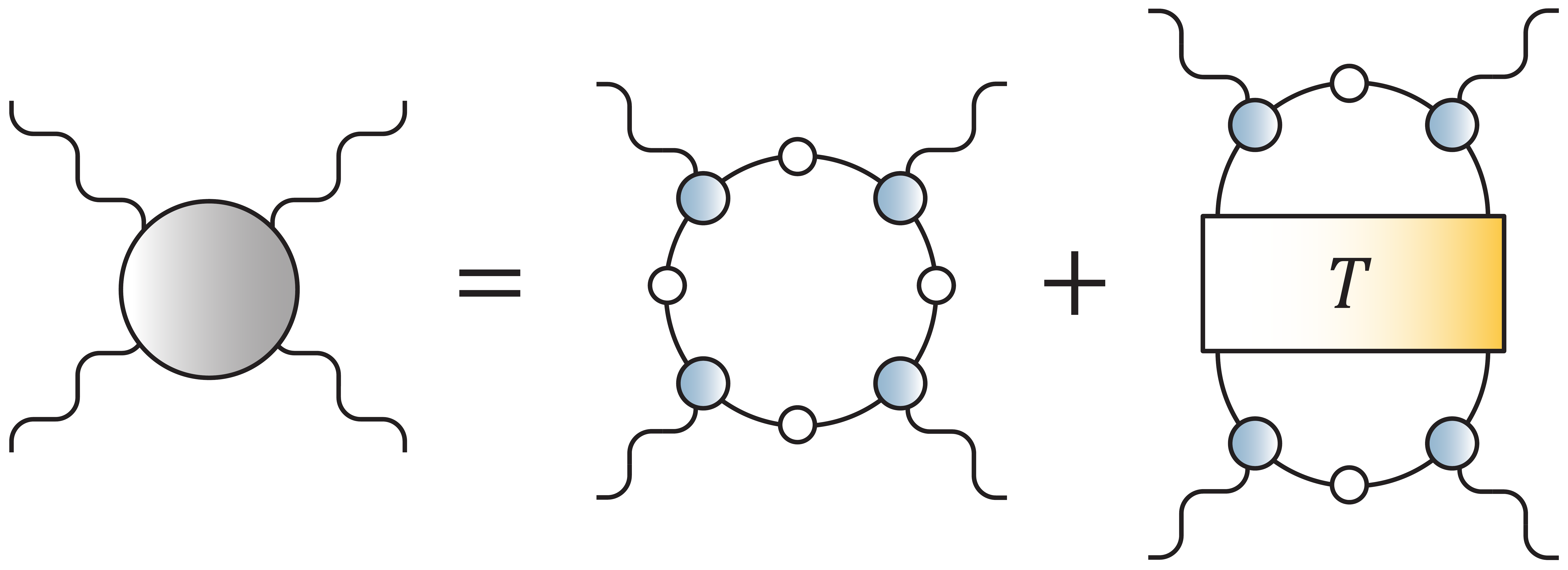}
                    \caption{Quark-level decomposition of the LbL amplitude in rainbow-ladder truncation.
                             Prefactors and permutations are omitted.}\label{fig:lbl-amplitude}
        \end{figure}

      Unfortunately, the remaining parts of the Ball-Chiu vertex including $\Delta_A$ and $\Delta_B$
      lead to an erratic numerical behavior which has not yet been resolved.
      Because the DSE result was so far obtained by neglecting these parts, it clearly cannot be gauge invariant.
      Can it be considered reliable under these circumstances?
      It was noted in Ref.~\cite{Goecke:2012qm} that changing the gauge of the photon propagator has only little effect
      on the quoted value; i.e., although gauge invariance is broken, the gauge artifacts appear to be negligible.
      We will return to this point below and study the magnitude of potential gauge artifacts in detail.

     \subsection{Towards a complete expression}

      We should stress in this context that there is \textit{a priori} no  reason why even the full quark loop \textit{should} be gauge invariant.
      Fig.~\ref{fig:lbl-amplitude} shows the `rainbow-ladder' decomposition of the LbL amplitude
     at the quark level that follows from electromagnetic gauge invariance~\cite{Eichmann:2011ec,Goecke:2012qm}.
     It contains the quark loop plus a graph where the connected quark-antiquark four-point function (the `quark scattering matrix')
     appears.
     Only the sum has to be gauge invariant, but not necessarily the individual contributions.
     The scattering matrix contains all intermediate meson poles and thereby reproduces the gauge-invariant hadronic exchange terms in Fig.~\ref{fig:lbl-amplitude-models}.
     In addition, it provides a natural offshell extension for those diagrams,
     which also serve to cancel potential gauge artifacts coming from the quark loop.
     Observe that the quark loop is \textit{necessary} to preserve gauge invariance,
     i.e., there is in fact no double-counting.

     Fig.~\ref{fig:lbl-amplitude} is not yet the complete expression for the LbL amplitude
     but only gauge invariant in a rainbow-ladder truncation, where the quark-quark interaction reduces to a gluon exchange.
     The scattering matrix is then a sum of gluon ladders; if the gluon shrinks to a point, it becomes
     the bubble sum from the ENJL model. Rainbow-ladder results are available for a range of hadron properties:
     spectra of light and heavy mesons~\cite{Maris:1999nt,Holl:2004fr,Blank:2011ha,Fischer:2014xha},
     their decay constants, form factors and other quantities~\cite{Maris:1997hd,Maris:1999bh,Maris:2002mz,Maris:2003vk,Maris:2005tt,Chang:2011vu},
     baryon spectra~\cite{Eichmann:2011vu,SanchisAlepuz:2011jn,Sanchis-Alepuz:2014sca}
     and a variety of nucleon and $\Delta$ elastic and transition form factors~\cite{Eichmann:2011vu,Eichmann:2011pv,Eichmann:2011aa,Sanchis-Alepuz:2013iia}.
     The analogue of Fig.~\ref{fig:lbl-amplitude} for $\pi\pi$ scattering was calculated
     in Ref.~\cite{Cotanch:2002vj}. Many of these observables have been obtained to good accuracy; for example,
     the value for $a_\mu^\text{(PS)} = (8.1 \pm 1.2) \times 10^{-10}$ from  the pseudoscalar exchange-diagram in Fig.~\ref{fig:lbl-amplitude-models}, where the rainbow-ladder solutions for the
     $\pi,\eta\to\gamma^\ast\gamma^\ast$ form factors are employed, agrees well with the model results quoted earlier~\cite{Goecke:2010if}.
     The systems where rainbow-ladder \textit{fails} are also known,
     and calculations beyond rainbow-ladder are well underway~\cite{Bender:1996bb,Watson:2004kd,Bhagwat:2004hn,Fischer:2005en,Matevosyan:2006bk,Fischer:2007ze,Fischer:2008wy,Alkofer:2008tt,Chang:2009zb,Chang:2010hb,Fischer:2009jm,Heupel:2014ina,Eichmann:2014xya,Williams:2014iea,Sanchis-Alepuz:2014wea}.
     So far, they have been mostly applied to mass spectra because
     the extension to structure properties (form factors etc.) is considerably more difficult.

     In the case of LbL scattering, Fig.~\ref{fig:lbl-amplitude} would pick up further diagrams where the photons also couple
     to the quark-antiquark kernel. One can show that a subset of these graphs reproduces the pion loop,
     which is the only diagram  in Fig.~\ref{fig:lbl-amplitude-models} that cannot be realized with the decomposition in Fig.~\ref{fig:lbl-amplitude}.
     Effects beyond rainbow-ladder will then primarily add to the `pion cloud', which is a situation very similar to the observables mentioned above.
     However, from the magnitude of the pion-loop contribution to $a_\mu$ one can infer that such effects will be rather small in the present case.
     A complete rainbow-ladder calculation of Fig.~\ref{fig:lbl-amplitude} should then provide a rough, but gauge-invariant, estimate for the value of $a_\mu$
     whose efficacy is tested by other hadronic processes such as those mentioned before.

     We finally note that both graphs in Fig.~\ref{fig:lbl-amplitude} can be absorbed into a one-loop diagram that contains
     two quark-photon vertices combined with a quark \textit{two-photon} vertex.
     The latter is the quark Compton vertex that is also responsible for the handbag-like diagrams in nucleon Compton scattering,
     together with the $t-$channel meson pole structure~\cite{Eichmann:2012mp}. Its Born terms are those that appear in the quark loop of the LbL amplitude,
     whereas the quark scattering matrix in the second graph constitutes the one-particle-irreducible part of the vertex.
     The structure is similar to the pion loop in Fig.~\ref{fig:lbl-amplitude-models} which is also not gauge invariant by itself,
     but gauge invariance can be restored by adding diagrams that contain the pion's Compton scattering vertex~\cite{Kinoshita:1984it,Engel:2012xb}.

         \begin{figure}[t]
                    \includegraphics[scale=0.15]{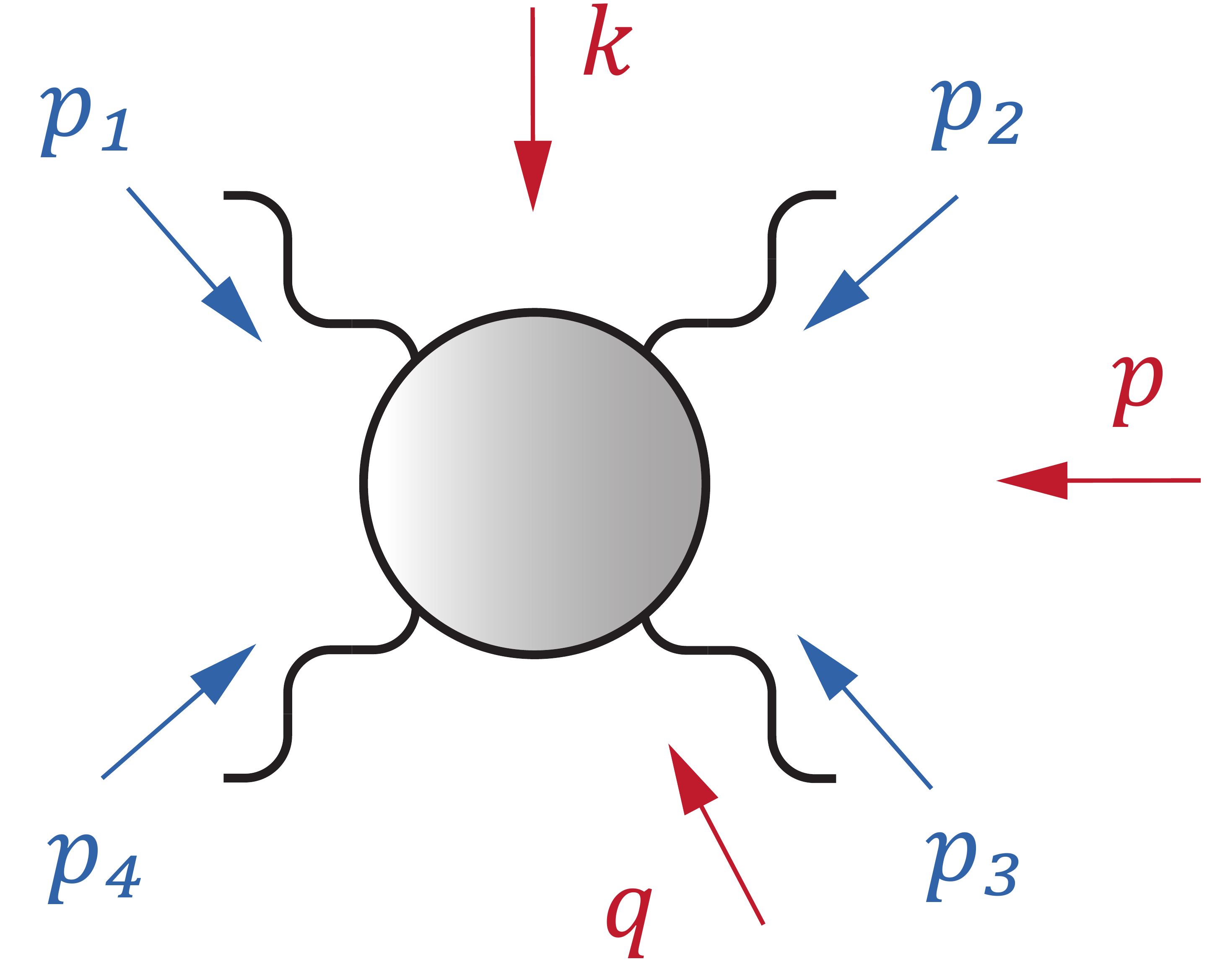}
                    \caption{ Kinematics in the photon four-point function.
                                            }\label{fig:kinematics}
        \end{figure}

        \begin{figure}[t]
          \includegraphics[scale=0.16]{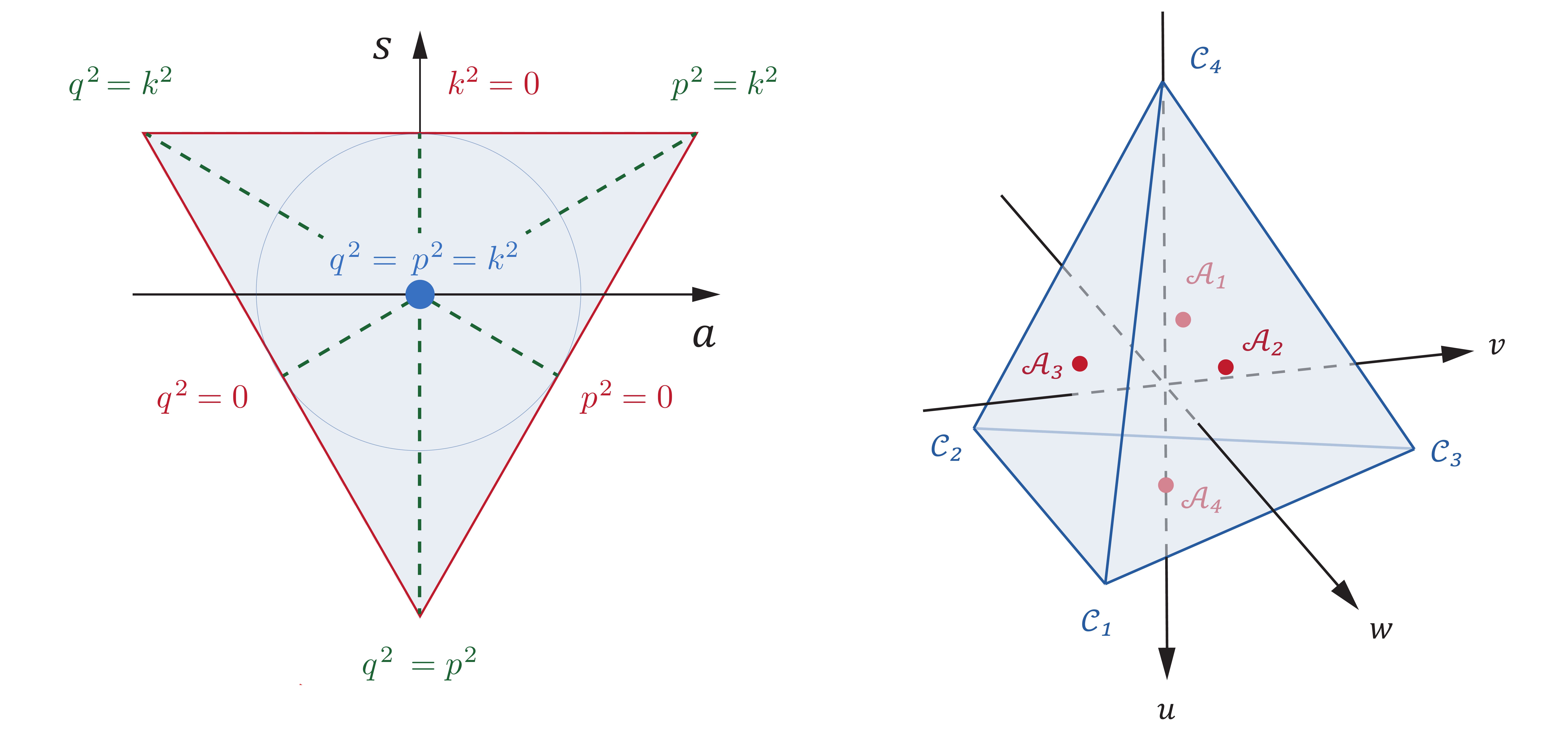}
          \caption{ Doublet triangle in the $(a,s)$ plane
                    and triplet tetrahedron in $(u,v,w)$ space, both at fixed $\mS_0$.
                    }\label{fig:phasespace}
        \end{figure}

\subsection{Structure of the light-by-light amplitude}

        Let us take a step back and study
        the structure of the LbL amplitude itself, without reference to any particular approach to calculate it.
        This should help to isolate the relevant kinematic regions
        that contribute to the g-2 integral, the importance of individual tensor structures, etc.

        The generic form of the LbL amplitude is given by
         \begin{equation}\label{4PA-decomposition}
             M^{\mu\nu\rho\sigma}(p,q,k) = \sum_{i=1}^{136} f_i(\Omega)\,\tau_i^{\mu\nu\rho\sigma}(p,q,k)\,,
         \end{equation}
         where $p$, $q$ and $k$ are three independent momenta upon which the amplitude depends.
         A convenient choice are the $s-$, $u-$ and $t-$channel momenta:
         \begin{equation}\label{Mandelstam-momenta}
         \begin{split}
             p &= p_2+p_3 = -p_1-p_4\,, \\
             q &= p_3+p_1 = -p_2-p_4\,, \\
             k &= p_1+p_2 = -p_3-p_4\,,
         \end{split}
         \end{equation}
         with the incoming photon momenta denoted by $p_i$, see Fig.~\ref{fig:kinematics}.
          $\Omega=\{ p^2,q^2,k^2,\omega_1,\omega_2,\omega_3\}$ comprises the six Lorentz invariants that constitute the phase space:
         from the momenta above one can form
         the three Mandelstam variables $p^2$, $q^2$, $k^2$ and the angular variables
         \begin{equation}\label{angular-variables}
             \omega_1=q\cdot k, \qquad
             \omega_2=p\cdot k, \qquad
             \omega_3=p\cdot q\,.
         \end{equation}
         The amplitude depends on 136 linearly independent tensor structures which we will discuss further below.

        Bose symmetry is a powerful assistant for investigating the LbL amplitude because it must be
        symmetric under any exchange of indices and corresponding momenta (e.g. $\mu\leftrightarrow\nu$, $p_1\leftrightarrow p_2$).
        One can cast both the dressing functions and the tensor structures in Eq.~\eqref{4PA-decomposition} into multiplets
        of the permutation group $\mathds{S}^4$ and combine them so that the full amplitude becomes a singlet.
        A function of four variables has 24 permutations; they can be rearranged into multiplets
        that transform under the irreducible representations of $\mathds{S}^4$. In terms of Young diagrams
        one has the following possibilities:
        \begin{equation}\label{S4-multiplets}
             \begin{array}{c @{\qquad} c @{\qquad} c @{\qquad\;\;} c @{\qquad\;\;} c}
             \mS & \mT^+_i & \mD_j & \mT^-_i & \mA \\[2mm]
             \tiny \yng(4) &
             \tiny\yng(3,1) &
             \tiny\yng(2,2) &
             \tiny\yng(2,1,1) &
             \tiny\yng(1,1,1,1)
             \end{array}
        \end{equation}
        $\mS$ and $\mA$ are completely symmetric or antisymmetric singlets.
        The doublets $\mD_j$ ($j=1,2$) form a two-dimensional irreducible subspace,
        and the triplets $\mT^+_i$ and antitriplets $\mT^-_i$ ($i=1,2,3$) form two three-dimensional subspaces.
        Their elements can be arranged in column vectors.

         For example, one can collect the six Lorentz invariants into a symmetric singlet, a doublet and a triplet:
        \renewcommand{\arraystretch}{1.0}
         \begin{align}\label{multiplets-LIs}
             \mS_0 &= \frac{p^2+q^2+k^2}{4}  \,, \nonumber \\[1mm]
             \mD_0 &= \mS_0 \left[ \begin{array}{c} a \\ s \end{array}\right] = \frac{1}{4} \left[ \begin{array}{c} \sqrt{3}\,(q^2-p^2) \\ p^2+q^2-2k^2 \end{array}\right], \\[2mm]
             \mT_0 &= \mS_0  \left[ \begin{array}{c} u \\ v \\ w \end{array}\right]
                    = \frac{1}{4}\left[ \begin{array}{c} -2\,(\omega_1+\omega_2+\omega_3) \\ -\sqrt{2}\,(\omega_1+\omega_2-2\omega_3) \\ \sqrt{6}\,(\omega_1-\omega_2)  \end{array}\right]  . \nonumber
         \end{align}
        We pulled out factors of $\mS_0$ to remove the mass dimension of the doublet and triplet variables ($a,s$ and $u,v,w$), so that only
        $\mS_0 \in \mathds{R}_+$ carries a dimension. The latter is a singlet because it can be written as
         \begin{equation}\label{multiplets-LIs}
             \mS_0 =  \frac{x_1+x_2+x_3+x_4}{4}\,,
         \end{equation}
         where   $x_i=p_i^2$ are the photon virtualities.

        The doublet phase space is the Mandelstam plane spanned by the variables $a$ and $s$ . It encodes the relations between
        $p^2$, $q^2$ and $k^2$ which are illustrated in Fig.~\ref{fig:phasespace}.
        Since the three momenta are independent and the variables $p^2$, $q^2$ and $k^2$ can take any value $\in \mathds{R}_+$,
        the spacelike region that contributes to the g-2 integral forms the interior of an equilateral triangle with side length $2\sqrt{3}$, enclosed by the lines
        $p^2=0$, $q^2=0$ and $k^2=0$.
        The three corners
        \begin{equation} \renewcommand{\arraystretch}{1.0}
            \left[\begin{array}{c} a \\ s \end{array}\right] \; = \;
            \left[\begin{array}{c} 0 \\ -2 \end{array}\right], \quad
            \left[\begin{array}{c} \sqrt{3} \\ 1 \end{array}\right], \quad
            \left[\begin{array}{c} -\sqrt{3} \\ 1 \end{array}\right]
        \end{equation}
        correspond to the limits $q^2=p^2=0$, $p^2=k^2=0$ and $q^2=k^2=0$, respectively.
        The timelike pion, scalar or axialvector two-photon poles will then form further triangles
        that enclose the spacelike one, and their strength will influence the shape of the LbL dressing functions inside the spacelike region.

         To discuss the triplet phase space it is advantageous to express $\mT_0$
         through the photon virtualities $x_i=p_i^2$:
         \begin{equation}\label{multiplets-LIs}
             \mT_0 = \frac{1}{4}\left[ \begin{array}{c} x_1+x_2+x_3-3x_4 \\ -\sqrt{2}\,(x_1+x_2-2x_3) \\ \sqrt{6}\,(x_1-x_2) \end{array}\right].
         \end{equation}
         For $x_i\in \mathds{R}_+$
         the resulting phase space forms the tetrahedron shown in Fig.~\ref{fig:phasespace}.
         In analogy to the doublet, the faces of the tetrahedron are the limits of vanishing photon virtualities $x_i=0$.
         Since each quark-photon vertex contributes vector-meson poles starting at $p_i^2=-m_\rho^2$, these poles will form further
         tetrahedra that enclose the one in Fig.~\ref{fig:phasespace}.
         On the other hand, the constraint $p_1 + p_2+p_3+p_4=0$ entails that not the full tetrahedron is probed under the g-2 integral.
         Instead, for fixed singlet and doublet variables, the triplet phase space is a rather complicated geometric object bounded by the tetrahedron.

\begin{figure}[t]
         \includegraphics[scale=0.14]{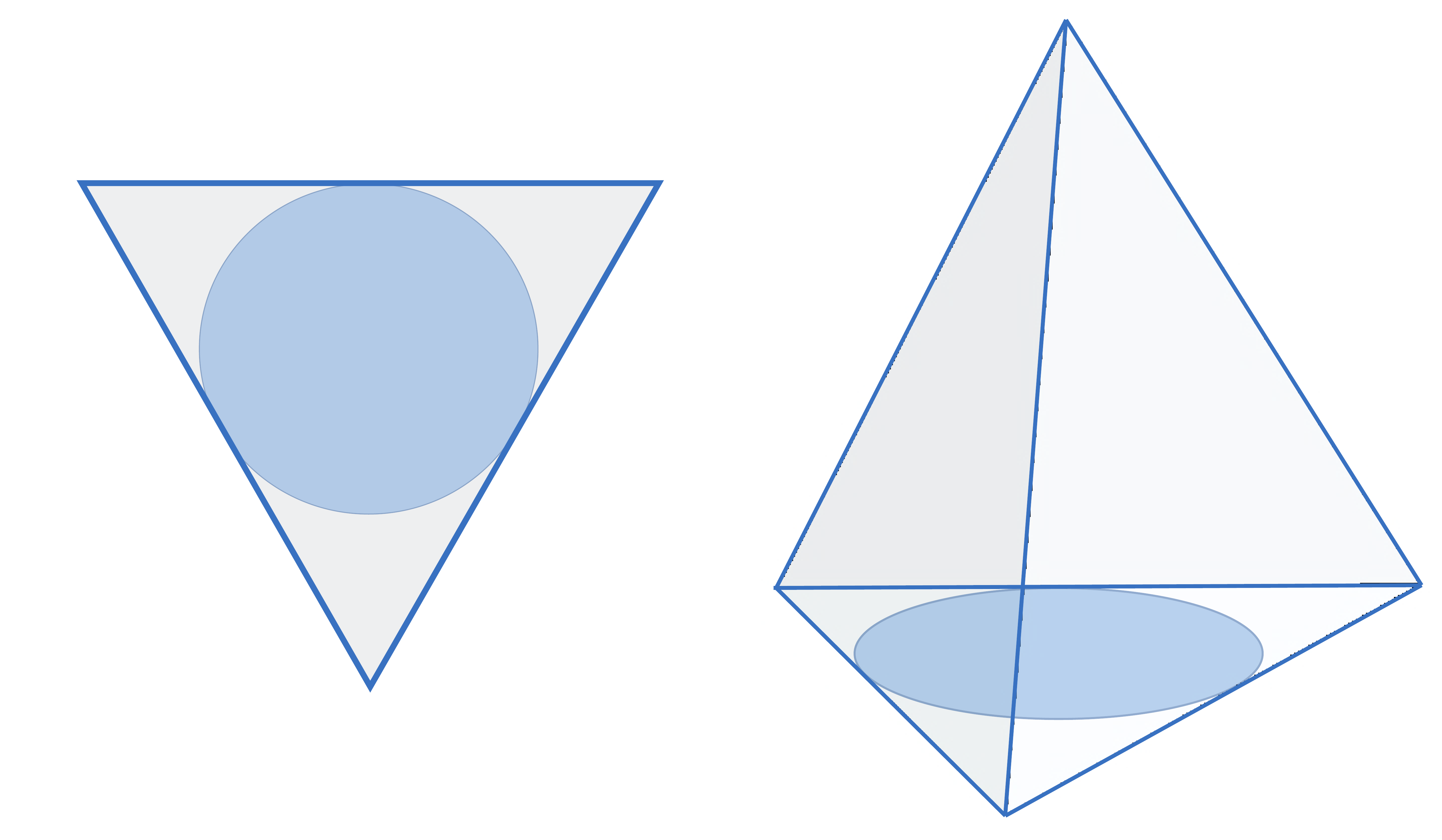}
         \caption{ Phase space that is relevant for $g-2$.}\label{fig:phasespace2}
       \end{figure}

        The kinematic limit that is relevant for the g-2 integration corresponds to one vanishing external photon momentum.
        For example, setting $p_4=0$ in Eq.~\eqref{Mandelstam-momenta} leads to $p+q+k=0$, which entails
        that the doublet and triplet variables are no longer independent but satisfy
         \begin{equation}\label{multiplets-LIs}
             \mT_0 = \mS_0  \left[ \begin{array}{c} 1 \\ -\sqrt{2}\,s \\ -\sqrt{2}\,a \end{array}\right], \quad a^2+s^2 \leq 1\,.
         \end{equation}
         The remaining phase space is visualized in Fig.~\ref{fig:phasespace2}: it is the unit circle within the triangle
         and the circle with radius $\sqrt{2}$ at the bottom of the tetrahedron.

    \renewcommand{\arraystretch}{1.5}

        \begin{table}[t]

                \begin{tabular}{  @{\;\;} l @{\;\;\;\;}     @{\;\;\;}l@{\;\;\;}   @{\;\;\;}l@{\;\;\;}   @{\;\;\;\;}l@{\;\;}      }

                           \noalign{\smallskip}
                           \hline

                    $n$  & \textbf{Seed}         &  \#       & \textbf{Multiplet}       \\    \noalign{\smallskip}\hline\noalign{\smallskip}

                    $0$               &$\delta^{\mu\nu} \delta^{\rho\sigma}$             &  $3$     & $\mS$,  $\mD_1$    \smallskip \\
                                      \noalign{\smallskip}\hline\noalign{\smallskip}

                    $2$               & $\delta^{\mu\nu}\,k^\rho\,k^\sigma$               &  $6$     & $\mS$, $\mD_1$, $\mT_1^+$     \\
                                      & $\delta^{\mu\nu}\,p^\rho\,p^\sigma$               &  $12$    & $\mS$, $\mD_1$, $\mD_2$, $\mT_1^\pm$, $\mA$     \\
                                      & $\delta^{\mu\nu}\,p^\rho\,q^\sigma$               &  $12$    & $\mS$, $\mD_1$, $\mT_1^+$, $\mT_2^\pm$     \\
                                      & $\delta^{\mu\nu}\,p^\rho\,k^\sigma$               &  $24$    & $\mS$, $\mD_1$, $\mD_2$, $\mT_1^\pm$, $\mT_2^\pm$, $\mT_3^\pm$, $\mA$ \smallskip  \\
                                      \noalign{\smallskip}\hline\noalign{\smallskip}

                    $4$               & $p^\mu\,p^\nu\,p^\rho\,p^\sigma$             &  $3$     & $\mS$, $\mD_1$    \\
                                      & $p^\mu\,p^\nu\,q^\rho\,q^\sigma$             &  $6$     & $\mS$, $\mD_1$, $\mT_1^-$   \\
                                      & $p^\mu\,p^\nu\,k^\rho\,k^\sigma$             &  $10$    & $\mS$, $\big(\mD_1$,\big) $\mD_2$, $\mT_1^\pm$, $\mA$     \\
                                      & $p^\mu\,q^\nu\,k^\rho\,k^\sigma$             &  $12$    & $\mS$, $\mD_1$, $\mT_1^+$, $\mT_2^\pm$  \\
                                      & $p^\mu\,p^\nu\,p^\rho\,k^\sigma$             &  $24$    & $\mS$, $\mD_1$, $\mD_2$, $\mT_1^\pm$, $\mT_2^\pm$, $\mT_3^\pm$, $\mA$     \\
                                      & $p^\mu\,p^\nu\,q^\rho\,k^\sigma$             &  $24$    & $\mS$, $\mD_1$, $\mD_2$, $\mT_1^\pm$, $\mT_2^\pm$, $\mT_3^\pm$, $\mA$     \smallskip \\
                                      \noalign{\smallskip}\hline\noalign{\smallskip}

                \end{tabular}

               \caption{         136-dimensional tensor basis for the LbL amplitude.
                                 $n$ denotes the mass dimension, and
                                 the overcounted doublet is shown in brackets.}
               \label{tab:generic-basis}

        \end{table}

         \renewcommand{\arraystretch}{1.0}

         The permutation-group arrangement has another practical advantage.
         If one succeeds in casting the tensor structures in permutation-group singlets,
         then the dressing functions must be also singlets, so they can only depend on symmetric variables.
         $\mS_0$ is symmetric, but all further variables emerge
         from products of $\mD_0$ and $\mT_0$ with higher momentum powers, for example
        \begin{equation}
        \begin{split}
             \mD_0\cdot\mD_0 &= \frac{3}{8}\,(p^4+q^4+k^4) - 2\,\mS_0^2\,, \\
             \mT_0\cdot\mT_0 &= \frac{3}{4}\,(\omega_1^2+\omega_2^2+\omega_3^2),
        \end{split}
        \end{equation}
         and also more complicated combinations.
         It is then conceivable that the dependence in the `angular' variables contained in
         $\mD_0$ and $\mT_0$ is small, similar to other systems where permutation-group arguments apply:
         the three-gluon vertex~\cite{Eichmann:2014xya}, the quark-photon vertex~\cite{Eichmann:2014qva}, etc.,
         so that the main momentum evolution comes from $\mS_0$.

         Table~\ref{tab:generic-basis} collects all linearly independent basis elements
         that can be constructed from the elementary permutation-group `seeds'
         \begin{equation}
             \delta^{\mu\nu} \delta^{\rho\sigma}, \quad
             \delta^{\mu\nu}\,c^\rho\,d^\sigma, \quad
             a^\mu\,b^\nu\,c^\rho\,d^\sigma,
         \end{equation}
         where $a,b,c,d \in \{ p,q,k\}$. In principle, each seed can produce
         any of the multiplets in Eq.~\eqref{S4-multiplets}:
         one singlet $\mS$, one antisinglet $\mA$, two doublets $\mD_j$, etc.,
         although many of them will vanish. This leads to the 138 elements in the table.
         Only 136 of those are, however, truly linearly independent: this is a consequence of the
         dimensionality of spacetime and affects all higher $n-$point functions.
         A similar restriction occurs in nucleon Compton scattering~\cite{Tarrach:1975tu}.
         Therefore, one must remove two elements from the basis in such a way that no
         kinematic singularities are introduced in the remaining dressing functions.
         This is a rather nontrivial task,
         and it turns out that the only safe choice is to remove the doublet $\mD_1$ for the
         seed element $p^\mu\,p^\nu\,k^\rho\,k^\sigma$ (shown in brackets in the table).

\begin{figure}[t]
           \includegraphics[scale=0.105]{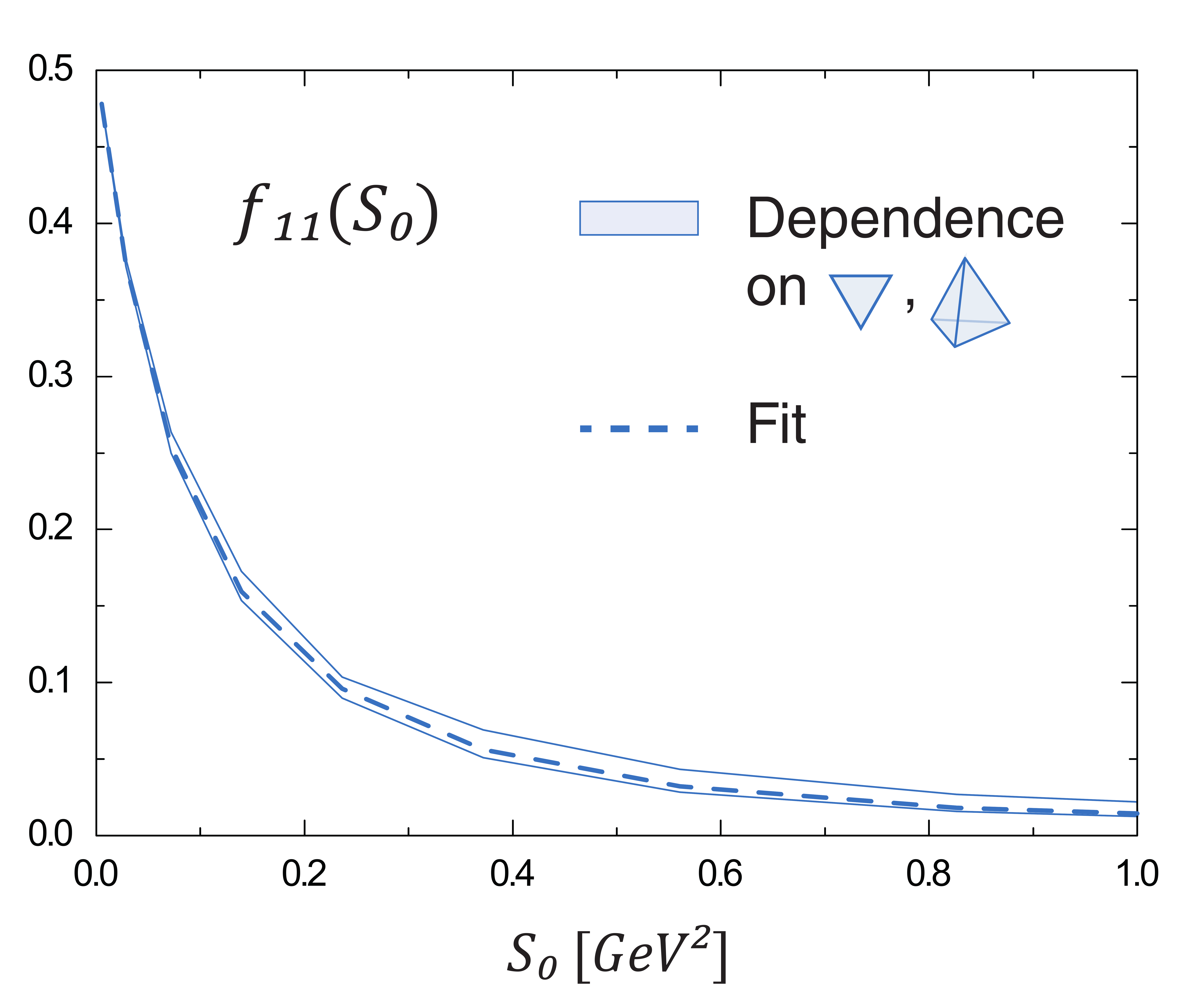}
           \caption{ Dressing function of the lowest-dimensional singlet tensor structure obtained from $p^\mu\,p^\nu\,q^\rho\,k^\sigma$
                     in Table~\ref{tab:generic-basis}, plotted as a function of $\mS_0$.
                     The band contains the full angular dependence in the doublet and triplet variables.}\label{fig:example}
         \end{figure}

         Going further, one can form 136 singlet basis elements from the multiplets in Table~\ref{tab:generic-basis}
         via appropriate combinations with the Lorentz-invariant multiplets $\mD_0$ and $\mT_0$.
         The resulting singlet form factors in that basis behave indeed as anticipated:
         they are essentially only functions of $\mS_0$, whereas the angular dependencies are almost negligible.
         We illustrate one exemplary result in Fig.~\ref{fig:example}:
         it is the dressing function of the lowest-dimensional
         singlet tensor structure obtained from $p^\mu\,p^\nu\,q^\rho\,k^\sigma$.
         The figure highlights the advantages of a symmetric tensor basis:
         it is sufficient to calculate the quark-loop integral in Fig.~\ref{fig:lbl-amplitude}
         on a \textit{sparse} grid of the six-dimensional phase space.
          The resulting dressing functions will look the same
          because their angular dependencies are small,
         and they can be fitted by functions that depend on $\mS_0$ only.
         Hence, the LbL amplitude can be well represented by 136 functions that depend only on \textit{one} variable.

\subsection{Gauge invariance}

         The basis in Table~\ref{tab:generic-basis} is not the most practical representation of the LbL amplitude because
         it does not yet implement the constraints from electromagnetic gauge invariance:
         \begin{equation}
             p_1^\mu \, M^{\mu\nu\rho\sigma} = 0, \quad \dots \quad
             p_4^\sigma \, M^{\mu\nu\rho\sigma} = 0\,.
         \end{equation}
         This will reduce the basis to a subset of 41 transverse tensors (as before, there are 43 possibilities but two of them are linearly dependent).
         Transversality and analyticity require these tensors to be proportional to at least four powers in the photon momenta.
         In principle one has to work out the Bose-symmetric condition
         \begin{equation}\label{transversality-1}
             T^{\mu\alpha}_{p_1}\,T^{\nu\beta}_{p_2}\,T^{\rho\gamma}_{p_3}\,T^{\sigma\delta}_{p_4}\,M^{\alpha\beta\gamma\delta} \stackrel{!}{=} M^{\mu\nu\rho\sigma}\,,
         \end{equation}
         where the $T_{p_i}^{\mu\nu}$ are the transverse projectors from Eq.~\eqref{trans-proj-0}.
         This leads to relations between the dressing functions;
         if we denote the independent functions by $f_i$ and the dependent ones by $g_j$, they take the form
         \begin{equation}
         \begin{split}
             g_1 &= g_1(f_1, \dots f_{41})\,, \\
                 &\vdots \\
             g_{95} &= g_{95}(f_1, \dots f_{41})\,.
         \end{split}
         \end{equation}
         They must be solved so that (a) no kinematic singularities are introduced in the process, i.e., all $g_j$ remain regular, and
         (b) only the minimal number of $g_j$ acquires a higher mass dimension by being proportional to any of the singlet variables $\mS_0$, $\mD_0\cdot\mD_0$, etc.
         (assuming that we started from a symmetric tensor basis where all dressing functions are singlets.)

         \begin{figure}[t]
           \includegraphics[scale=0.30]{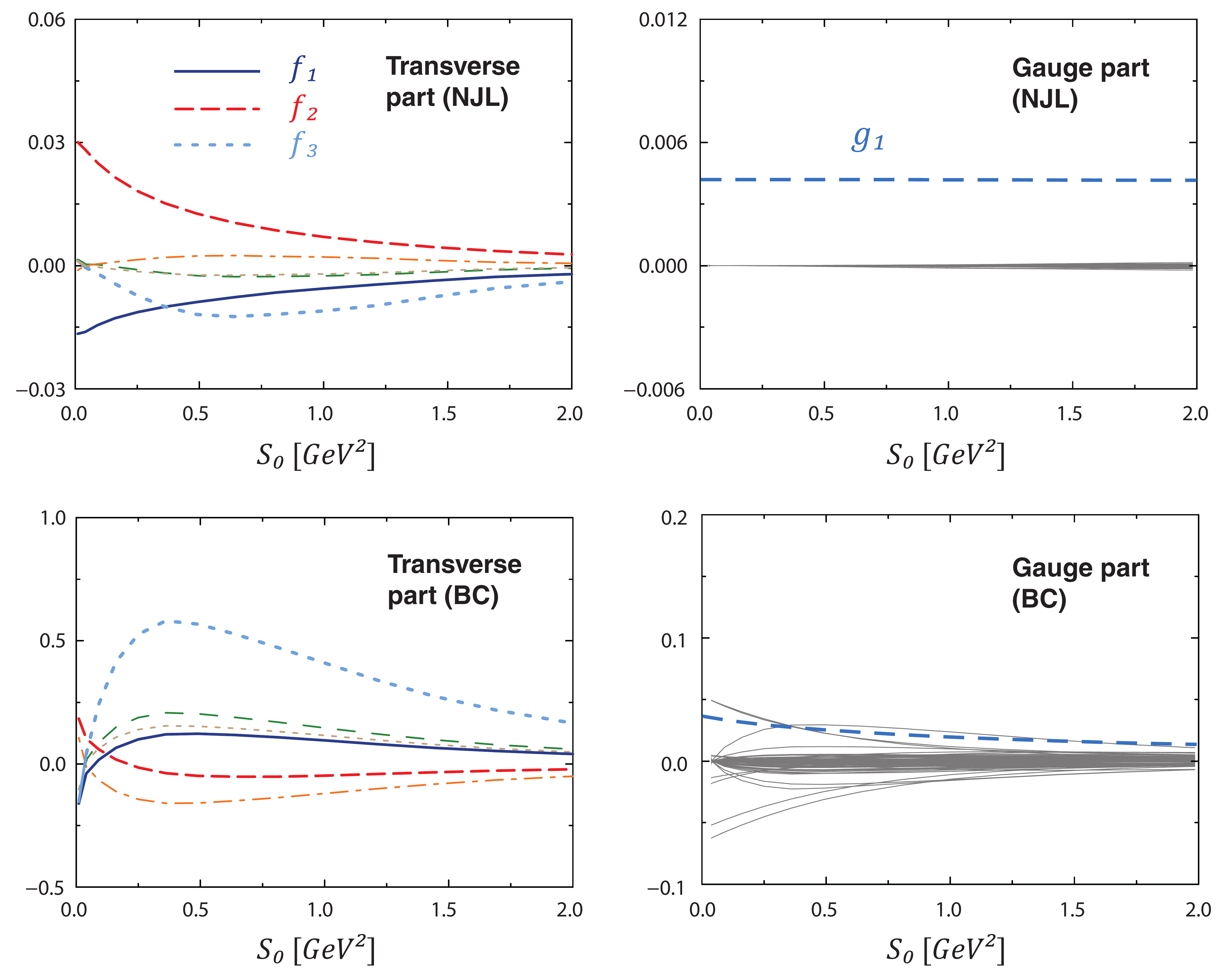}
           \caption{Dressing functions of the LbL amplitude obtained from a constituent-quark loop (\textit{upper panels})
                    and with a dressed quark loop employing a Ball-Chiu vertex (\textit{lower panels}). The legends only enumerate the dressing functions $f_{1,2,3}$ and $g_1$ that are also discussed in the text.}\label{fig:results}
         \end{figure}

         The resulting amplitude takes the form
         \begin{equation}\label{amp-1}
             M^{\mu\nu\rho\sigma} = \sum_{i=1}^{41} f_i\,\tau_{\perp i}^{\mu\nu\rho\sigma} + \sum_{j=1}^{95} g_j\,\tau_j^{\mu\nu\rho\sigma}.
         \end{equation}
         The first term is the transverse part of the amplitude, with transverse tensors $\tau_{\perp i}^{\mu\nu\rho\sigma}$ that have mass dimension $4,6,8\dots$,
         and dressing functions $f_i$ that become constant in any kinematic limit.
         The second term constitutes the `gauge part', which is neither longitudinal nor transverse, and must vanish if the amplitude is gauge invariant.
         (It cannot be chosen purely longitudinal because that would again introduce kinematic singularities.)
         The fact that the $\tau_j$ remain with mass dimension $0, 2, 4, \dots$
         is the reason why  violating gauge invariance can have severe consequences in practice.
         With another transverse projection of Eq.~\eqref{amp-1} everything collapses into the transverse part.
         If the dressing functions $g_j$ are nonzero, they will introduce artificial singularities with momentum powers $-4, -2$, etc. into the $f_i$.

         The analogous steps~(\ref{transversality-1}--\ref{amp-1}) for the nucleon's Compton scattering amplitude have been worked out by Tarrach~\cite{Tarrach:1975tu}.
         For simpler systems like the quark-photon vertex they are straightforward to solve~\cite{Kizilersu:1995iz,Eichmann:2012mp};
         in that case the gauge part is the Ball-Chiu vertex in Eq.~\eqref{vertex:BC}
         which remains nonzero because the right-hand side of Eq.~\eqref{qpv-wti} does not vanish, i.e., the vertex is not gauge invariant.
         By constrast, the situation for the LbL amplitude is much more complicated because there are many possible kinematic limits and the $g_j$
         must be regular in any of them.

         An alternative way to approach the problem is to  construct tensors with lowest possible mass dimensions that are automatically free of kinematic
         singularities. For example, one can employ
         \begin{equation}
         \begin{split}
             t^{\mu\nu}_{ij} &= p_i\cdot p_j\,\delta^{\mu\nu} - p_j^\mu \,p_i^\nu, \\
             \varepsilon^{\mu\nu}_{ij} &= \varepsilon^{\mu\nu\alpha\beta}\,p_i^\alpha\,p_j^\beta
         \end{split}
         \end{equation}
         as the building blocks for the construction of such tensors~\cite{Eichmann:2012mp}.
         $t^{\mu\nu}_{ij}$ is transverse with respect to $p_i^\mu$ and $p_j^\nu$, and $\varepsilon^{\mu\nu}_{ij}$ is transverse to both momenta.
         Some possible tensor structures for the LbL amplitude are then the following:
         \begin{equation}
             t^{\mu\nu}_{12}\,t^{\rho\sigma}_{34}\,, \quad
             \varepsilon^{\mu\nu}_{12}\,\varepsilon^{\rho\sigma}_{34}\,, \quad
             t^{\mu\alpha}_{11}\,t^{\alpha\nu}_{22}\,t^{\rho\sigma}_{34}\,, \quad \text{etc.}
         \end{equation}
         They can be used as permutation-group seeds in analogy to Table~\ref{tab:generic-basis}.
         The lowest dimension-four singlets are the tensors
         \begin{align}
             \tau_{\perp 1}^{\mu\nu\rho\sigma} &= t^{\mu\nu}_{12}\,t^{\rho\sigma}_{34} + t^{\nu\rho}_{23}\,t^{\mu\sigma}_{14} + t^{\rho\mu}_{31}\,t^{\nu\sigma}_{24}\,, \label{sc-sc} \\
             \tau_{\perp 2}^{\mu\nu\rho\sigma} &= \varepsilon^{\mu\nu}_{12}\,\varepsilon^{\rho\sigma}_{34} + \varepsilon^{\nu\rho}_{23}\,\varepsilon^{\mu\sigma}_{14} + \varepsilon^{\rho\mu}_{31}\,\varepsilon^{\nu\sigma}_{24}\,, \label{ps-ps}
         \end{align}
         whereas the others have a higher mass dimension.
         The construction elements $t^{\mu\nu}_{12}$ and $t^{\mu\alpha}_{11}\,t^{\alpha\nu}_{22}$ are the two form factors of a scalar two-photon current, and
         $\varepsilon^{\mu\nu}_{12}$ is that of a pseudoscalar ($\pi\to\gamma\gamma$) current. Hence, these tensors already reflect the pole
         structure of the LbL amplitude: the scalar pole will appear in $f_1$ attached to Eq.~\eqref{sc-sc}, the pion pole in $f_2$, and so on.

         The construction of a basis completely free of kinematic singularities is still under development.
         In the meantime some useful observations can be made.
         In Fig.~\ref{fig:results} we plot the quark-loop result for the dressing functions of the LbL amplitude.
         That is, we calculate the quark loop and project the full amplitude onto a particular tensor basis of the form given in Eq.~\eqref{amp-1}.
         The top panels show the results from a constituent-quark loop (with $m=500$ MeV) and the bottom panels those from a dressed loop where dressed propagators and the full Ball-Chiu vertex were implemented.
         The left panels depict the six largest dressing functions from the transverse part (out of 41), and the right panels contain \textit{all} 95 dressing functions in the gauge part.
         In contrast to Fig.~\ref{fig:example} we do not show the full angular dependence but rather a specific kinematic point within the triangle and the tetrahedron, plotted as a function of $\mS_0$.

         From the upper left plot one infers that the two functions $f_1$ and $f_2$ from Eqs.~(\ref{sc-sc}--\ref{ps-ps}) are indeed among the dominant ones in the constituent-quark case.
         Within numerical accuracy, the dressing functions from the gauge part are all zero --- with the exception of $g_1$ which corresponds to the `tree-level' tensor structure
         \begin{equation}\label{gauge-part-t1}
             \tau_1^{\mu\nu\rho\sigma} = \delta^{\mu\nu}\,\delta^{\rho\sigma} + \delta^{\nu\rho}\,\delta^{\mu\sigma} + \delta^{\rho\mu}\,\delta^{\nu\sigma}\,
         \end{equation}
         and turns out to be a constant. The value of this constant can be obtained analytically using a Feynman parametrization, it is $g_1 = 1/(24 \pi^2)$.
         Hence, not even the constituent-quark loop is gauge invariant, but
         the problem is avoided in practical calculations of $a_\mu$ where one employs the identity
         \begin{equation}
             M^{\mu\nu\rho\sigma} = -p_4^\lambda\,\frac{d}{dp_4^\sigma}\,M^{\mu\nu\rho\lambda}\,,
         \end{equation}
         which holds if $M^{\mu\nu\rho\sigma}$ is transverse. A nonzero gauge part will drop out
         as long as it is constant, which is the case here: Eq.~\eqref{gauge-part-t1} is the only possible singlet tensor structure that is momentum-independent.

         Turning now to the dressed quark loop in the bottom panels, we find that both the magnitude and the relative strengths of the transverse dressing functions change dramatically.
         These are not necessarily genuine effects. On the one hand, we used a relatively large quark mass $m=500$ MeV for the NJL case and the $f_i$ are not dimensionless: $f_1$ and $f_2$ scale with $1/m^4$, etc.
         Halving the quark mass would increase their magnitude already by a factor 16. On the other hand,
         our transverse basis is not yet free of kinematic singularities. This can be already deduced from the fact
         that several dressing functions vanish with $\mS_0 \rightarrow 0$ instead of approaching constant, nonzero values.
         As another example, the magnitude of $f_3$ (whose tensor structure is an inconspicuous singlet constructed from one of the triplets) is due to a singularity which happens in a particular kinematic limit.
         A `minimal' basis should be free of such problems and the dressing functions plotted over the full angular domain should rather resemble those in Fig.~\ref{fig:example}.

         Note also that the gauge part is no longer zero: some of the $g_i$ are now comparable in size to $g_1$, which has picked up a momentum dependence.
         They are still small compared to the transverse dressing functions (we zoomed into the right panels with a factor of 5 compared to the left panels),
         and with different tensor decompositions many of them could drop out.
         Nevertheless it is clear that such a non-zero gauge part will complicate or even invalidate a clean extraction of $a_\mu$.
         In principle, the decomposition of Eq.~\eqref{amp-1} can be used as a 'filter': keeping only the transverse part should produce sensible results
         as long as one has a tensor basis that is kinematically safe.
         In any case, a non-transverse quark loop is not a problem \textit{per se} because all non-zero functions $g_i$ must be canceled
         by the second diagram in Fig.~\ref{fig:lbl-amplitude}, which contains the `offshell contributions' to the meson pole terms.

         We finally recall that we obtained the bottom panels of Fig.~\ref{fig:results} using the \textit{full} Ball-Chiu vertex,
         whereas the present DSE result for $a_\mu^\text{(QL)}$ from Ref.~\cite{Goecke:2012qm} includes only the dressing of the $\gamma^\mu$ part, together with the leading transverse component. The fact that
         its value was found to be stable under variations of the gauge parameter indicates that the non-zero gauge parts in Fig.~\ref{fig:results}  are mainly
         a consequence of the vertex dressings $\Delta_A$ and $\Delta_B$ in Eq.~\eqref{vertex:BC}.
         In turn, this implies that the current result for $a_\mu^\text{(QL)}$ is still largely unaffected by gauge artifacts and therefore reliable,
         although efforts to update its value are underway.

         \subsection{Summary}

         We summarized recent developments in the Dyson-Schwinger approach regarding the light-by-light scattering contribution to the muon anomalous magnetic moment.
         We discussed the structure of the light-by-light amplitude and the consequences of gauge invariance,
         and we identified the regions of the phase space that provide support for the g-2 integral. We emphasize again that a microscopic description of the amplitude
         accommodates all diagrams that have been studied so far with models: quark loops, hadronic exchanges including offshell effects, and in principle also pseudoscalar loops.
         An inclusion of the quark loop is in fact necessary to satisfy electromagnetic gauge invariance for the full amplitude
         if the quark propagator has a realistic momentum evolution.

         The current best value for the dressed quark loop from the Dyson-Schwinger approach is~\cite{Goecke:2012qm}
         \begin{equation}
            a_\mu^\text{(QL)} = 10.7(2) \times 10^{-10}\,.
         \end{equation}
         It is an intermediary result because it was obtained by omitting certain components of the quark-photon vertex.
         We are working to update this number by calculating the fully dressed
         quark loop, together with the diagram that involves the scattering matrix and is needed to restore gauge invariance.
         The latter should automatically reproduce the hadronic exchange terms from the quark level;
         nevertheless, a direct result for the sum of pseudoscalar ($\pi$, $\eta$, $\eta'$) exchange diagrams is also available~\cite{Goecke:2010if}:
         \begin{equation}
            a_\mu^\text{(PS)} = (8.1 \pm 1.2) \times 10^{-10}\,.
         \end{equation}
         Finally, these numbers should be complemented by the Dyson-Schwinger result for the hadronic vacuum polarization~\cite{Goecke:2011pe,Goecke:2013fpa}:
         \begin{equation}
            a_\mu^\text{(HVP)} = 676 \times 10^{-10}\,.
         \end{equation}

%%%%%%%%%%%%%%%%%%%%%%%%%%%%%%%%%%%%%%%%%%%%%%%%
%% BACKMATTER
%%%%%%%%%%%%%%%%%%%%%%%%%%%%%%%%%%%%%%%%%%%%%%%%

\begin{theacknowledgments}
  This work was supported by the German Science Foundation (DFG) under project number DFG TR-16, by the Helmholtz International Center for FAIR
  within the LOEWE program of the State of Hesse, and by the Helmholtzzentrum GSI.
\end{theacknowledgments}

\bibliographystyle{aipproc}   % if natbib is available

\bibliography{lit-g-2}

%%%%%%%%%%%%%%%%%%%%%%%%%%%%%%%%%%%%%%%%%%%
%% Just a reminder that you may have to run bibtex
%% All of it up to \end{document} can be removed
%% if you don't like the warning.
%%%%%%%%%%%%%%%%%%%%%%%%%%%%%%%%%%%%%%%%%%%

\IfFileExists{\jobname.bbl}{}
 {\typeout{}
  \typeout{******************************************}
  \typeout{** Please run "bibtex \jobname" to optain}
  \typeout{** the bibliography and then re-run LaTeX}
  \typeout{** twice to fix the references!}
  \typeout{******************************************}
  \typeout{}
 }

\end{document}